\documentclass[11pt]{amsart}
\usepackage{fancyhdr}
\usepackage{geometry}                % See geometry.pdf to learn the layout options. There are lots.
\usepackage{graphicx}
\usepackage{amsmath,amssymb}
\usepackage{appendix}
\usepackage{epstopdf}
\usepackage{url}
\usepackage[colorlinks]{hyperref}
\usepackage{color}
\definecolor{orange}{rgb}{1,0.5,0}
\definecolor{gray}{rgb}{0.5,0.5,0.5}
\definecolor{roetlich}{rgb}{1, .7, .7}
\definecolor{camel}{rgb}{0.76, 0.6, 0.42}
\definecolor{bronze}{rgb}{0.8, 0.5, 0.2}
\definecolor{britishracinggreen}{rgb}{0.0, 0.26, 0.15}
\newcommand{\beq}{\begin{equation}}
\newcommand{\eeq}{\end{equation}}

\newcommand{\fnequ}{{\rm neq/cm^2}}
\usepackage[]{lineno}
\title{Energy and time measurements with high-granular silicon devices}

\date{}                                           % Activate to display a given date or no date
\begin{document}

\maketitle

%\noindent
\begin{center}
MARCELLO MANNELLI$^1$, ROMAN P\"OSCHL$^2$\footnote{Corresponding author: poeschl@lal.in2p3.fr}, ABRAHAM SEIDEN$^3$\\
\vspace{0.2cm}
\begin{footnotesize}
\indent $^1$ {\it CERN, Organisation europ\'eenne pour la recherche nucl\'eaire, CH-1211 Gen\`eve 23, Switzerland}\\
\indent $^2$ {\it Laboratoire de l'Acc\'el\'erateur Lin\'eaire (LAL),\\ Centre Scientifique d'Orsay, B\^atiment 200\\ F-91898 Orsay CEDEX, France}\\
\indent $^3$ {\it UC Santa Cruz, 1156 High Street, Santa Cruz, CA 95064, United States of America}
\end{footnotesize}
\end{center}
\bigskip

\thispagestyle{fancy}

\newcommand{\ecm}{\mathrm{\sqrt{s}}}
\newcommand{\eplus}{\mathrm{e}^+}
\newcommand{\eminus}{\mathrm{e}^-}
\newcommand{\taul}{\mathrm{\tau}}
\newcommand{\epem}{\eplus\eminus}
\newcommand{\lplm}{l^+l^-}
\newcommand{\mpmm}{\mu^+\mu^-}
\newcommand{\tptm}{\tau^+\tau^-}
\newcommand{\eeX}{\epem X}
\newcommand{\mmX}{\mpmm X}
\newcommand{\ZH}{\mathrm{ZH}}
\newcommand{\Zzero}{\mathrm{Z}}
\newcommand{\Zo}{\mathrm{Z^0}}
\newcommand{\Wboson}{\mathrm{W}}
\newcommand{\WpWm}{\Wboson^+\Wboson^-}
\newcommand{\Higgs}{\mathrm{H}}
\newcommand{\quark}{\mathrm{q}}
\newcommand{\bottom}{\mathrm{b}}
\newcommand{\tpq}{\mathrm{t}}
\newcommand{\afb}{\mathrm{A_{FB}}}
\newcommand{\alr}{\mathrm{A_{LR}}}
\newcommand{\qq}{\mathrm{q}\overline{\mathrm{q}}}
\newcommand{\uubar}{\mathrm{u}\overline{\mathrm{u}}}
\newcommand{\ddbar}{\mathrm{d}\overline{\mathrm{d}}}
\newcommand{\ssbar}{\mathrm{s}\overline{\mathrm{s}}}
\newcommand{\ttbar}{\mathrm{t}\overline{\mathrm{t}}}
\newcommand{\GammaZ}{\Gamma_\Zzero}
\newcommand{\GammaW}{\Gamma_\Wboson}
\newcommand{\Gammat}{\Gamma_\tpq}
\newcommand{\mZ}{\mathrm{M_\Zzero}}
\newcommand{\mW}{\mathrm{M_\Wboson}}
\newcommand{\mH}{\mathrm{M_\Higgs}}
\newcommand{\mt}{\mathrm{M_\tpq}}
\newcommand{\Mdl}{\mathrm{M_{dl}}}
\newcommand{\Mrecoil}{\mathrm{M_{recoil}}}
\newcommand{\Ptdl}{\mathrm{P_{Tdl}}}
\newcommand{\Pdl}{\mathbf{P_{dl}}}
\newcommand{\roots}{\mathrm{\sqrt{s}}}
\newcommand{\invfb}{\mathrm{fb^{-1}}}
\newcommand{\invpb}{\mathrm{pb^{-1}}}
\newcommand{\TeV}{\mathrm{TeV}}
\newcommand{\GeV}{\mathrm{GeV}}
\newcommand{\MeV}{\mathrm{MeV}}
\newcommand{\GeVovc}{\mathrm{\GeV/c^2}}
\newcommand{\polRL}{\mathrm{e^-_R e^+_L}}
\newcommand{\polLR}{\mathrm{e^-_L e^+_R}}
\newcommand{\rmsn}{\mathrm{rms}_{90}}
\newcommand{\cthadl}{\mathrm{cos\theta_{dl}}}
\newcommand{\Ptbal}{\mathrm{\Delta P_{Tbal.}}}
\newcommand{\Ptgamma}{\mathrm{P_{T\gamma}}}
\newcommand{\Pt}{\mathrm{P_{T}}}
\newcommand{\Naddtks}{\mathrm{N_{add.TK}}}
\newcommand{\dthattk}{\mathrm{|\Delta \theta_{2tk}|}}
\newcommand{\dthamin}{\mathrm{|\Delta \theta_{min}|}}
\newcommand{\cthamiss}{\mathrm{|cos\theta_{missing}|}}

\hyphenation{brems-strah-lung}

\begin{abstract}
This note is a short summary of the workshop on {\em Energy and time measurements with high-granular silicon devices} that took place on the 13/6/16 and the 14/6/16 at DESY/Hamburg in the frame of the first AIDA-2020 Annual Meeting~\cite{bib:url-siws}. This note tries to put forward trends that could be spotted and to emphasise in particular open issues that were addressed by the speakers.

\end{abstract}

%\section{}
%\subsection{}

\section{Introduction}

Silicon is particularly well suited for the design of compact and highly segmented calorimeters. 
Calorimeters with silicon as active elements have a tradition that goes back to the LEP era. Small silicon-tungsten calorimeters (with a diameter of around 30\,cm) were used e.g. by the OPAL collaboration for luminosity measurements in the forward regions of the detector~\cite{Abbiendi:2002dx}. This "tradition" will be followed up by the luminosity calorimeter that is designed for future linear electron-positron colliders. A highly segmented calorimeter is also beneficial in more central regions of the detector. A first attempt was made by the ALEPH collaboration. The ALEPH detector featured an electromagnetic calorimeter using subdivided wire-proportional chambers~\cite{Buskulic:1994wz}. The vast development of silicon detectors for tracking with ever increasing surface renders it today possible to envisage silicon based  large surface central calorimeters for future experiments in particle physics. In the past 10-15 years the R\&D on these devices has been conducted within the CALICE collaboration~\cite{bib:url-calice} driven by the needs of future linear electron positron colliders. The success of this R\&D programme inspired the LHC collaboration to consider large scale calorimeters based on silicon for their high luminosity upgrades. The workshop brought together experts from the two communities to discuss the next steps towards the realisation of these detectors. 

\section{High radiation environment at HL-LHC}

\begin{figure}[h]
  {\centering
    \includegraphics[width=0.7\textwidth]{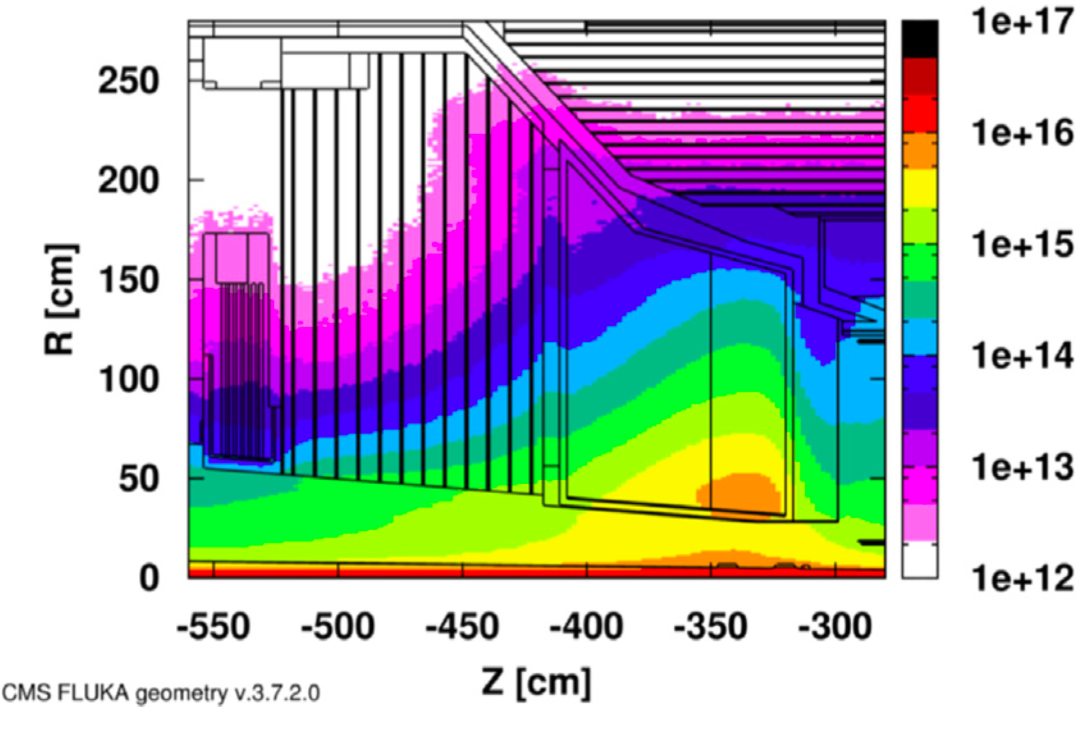}
    \caption{\sl  Equivalent 1\,MeV neutron fluence (right side y-axis) for Phase-II endcap calorimetry at HL-LHC. The figure is taken from~\cite{Curras:2212843}.}
    \label{fig:cms-fluence}
  }
\end{figure}

In particular the LHC experiments have to take into account high particle fluencies of up to $10^{16}\,\fnequ$, see Fig.~\ref{fig:cms-fluence}. 

\begin{figure}[h]
  {\centering
    \includegraphics[width=0.7\textwidth]{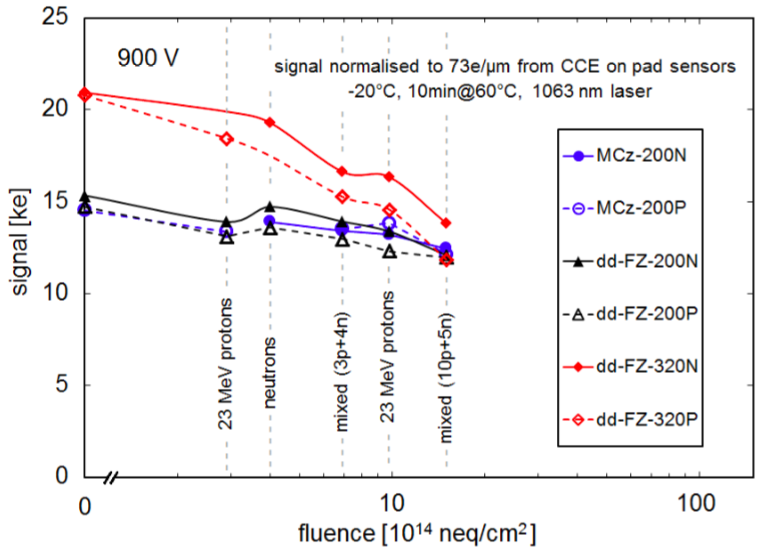}
    \caption{\sl  Signal variation of different Si sensor type after irradiation with increasing fluences by different sources. The figure is taken from~\cite{bib:phd-poehlsen}.}
    \label{fig:pn-irrad}
  }
\end{figure}

Different sensor thicknesses and different bias voltages were tested, see e.g.~\cite{Curras:2212843}. Thick sensors deliver higher signals but are at the same time more sensitive to radiation effects, i.e. the signal amplitude as a function of the fluency decreases more rapidly compared with thin sensors.
The choice of the type of the sensor plays an important role for the success of a detector. A basic  choice has to be made between n-in-p wafers. i.e. substrate is made from Si doped with a donator, called n-type hereafter, and p-in-n wafers, i.e. the substrate layer is doped with an acceptor, called p-type hereafter. The RD-50~\cite{bib:rd-50} and CMS collaborations~\cite{bib:cern-cms} have tested a number of n-type and p-type wafers. It turned out that after irradiation the signal generated by a MIP is smaller for p-type wafers than for n-type wafers, see Fig.~\ref{fig:pn-irrad}. 

It seems that this effect is larger in case of thick wafers and after irradiation. In particular the smaller signal after irradiation for thick wafers is also reported in~\cite{Curras:2212843}. It should be noted at this point that the described phenomena has so far been only observed for Si wafers  provided by HPK. Measurements on sensors by other producers, e.g. Micron, did not reveal such a behaviour~\cite{Kramberger:2010zza}. 

\section{Tools}
\begin{figure}[h]
  {\centering
    \includegraphics[width=0.4\textwidth]{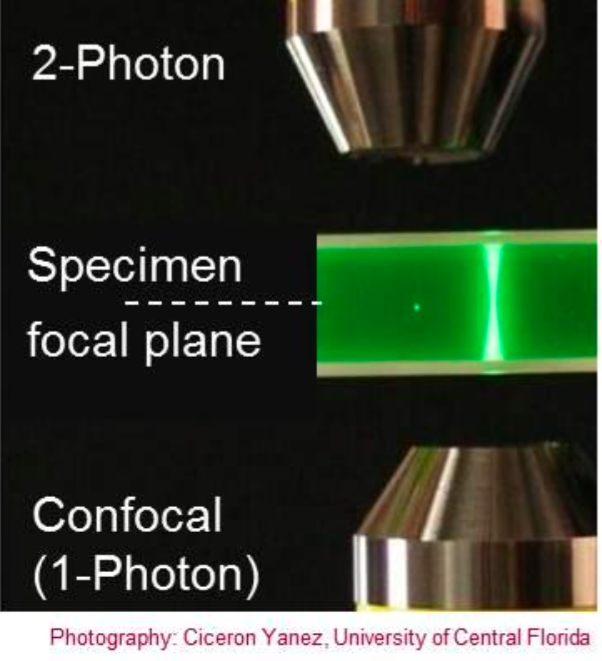}
    \caption{\sl  Comparison of illumination of probe with the two photon technique (left, little green spot) and the standard one-photon technique (right, green strip).}
    \label{fig:tpa-tct}
  }
\end{figure}

 A quite novel technology to characterise silicon sensors is based on two-photon absorption. In fact, two laser photons annihilate at a well defined point is space within the semiconductor material and produce an electron hole pair. The exact knowledge of the production point allows for example for a precise mapping of the charge collection efficiencies of silicon sensors. The advantage with respect to the "standard" one-photon absorption Transition Current Technique, in which the signal is produced across the full laser path, is obvious from Fig.~\ref{fig:tpa-tct}. 
At the workshop results for laser photons with a wavelength of 1300\,nm, i.e. the far infrared, were shown which corresponds to optimal wavelengths given in the literature.

The behaviour of silicon sensors is typically simulated using TCAD simulations. However, these simulations are often optimised for specific parameters. The RD-50 collaboration is leading an effort to understand the limitations of existing models and to obtain a consistent parameter set by performing global fits to different parameters (CV, IV and CCE)  within the TCAD simulator.

A new tool to learn on and to understand the behaviour of silicon wafers is the computer programme Weightfield 2.0~\cite{bib:weightfield}. Weightfield 2.0 allows for a detailed simulation of signal generation in silicon (and diamond sensors) as well as for a simulation of the subsequent signal shaping in the readout electronics. Applications of Weightfield 2.0 were shown at the workshop. These include for example the comparisons of pulse heights measured by CMS with those obtained by the Weightfield 2.0 simulation.

\section{Timing}

\begin{figure}[h]
  {\centering
    \includegraphics[width=0.6\textwidth]{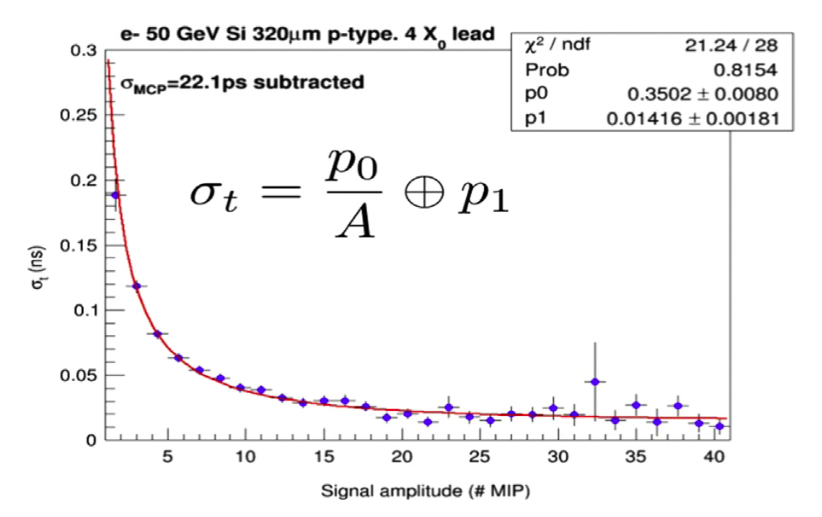}
    \caption{\sl  Time resolution as a function of the signal amplitude as measured in CMS beam test experiment. The figure is taken from~\cite{Curras:2212843}.}
    \label{fig:time-resol}
  }
\end{figure}

So far emphasis is put on the imaging capabilities of highly granular calorimeters due to the detailed spatial information available.
Progress in readout technology allows for adding a new dimension to the list of useful measurements, the time. The exploitation of this option is currently driven by the high interaction rate at the HL-LHC but it is most likely that any future facility, i.e. also linear electron-positron colliders will benefit from the expected progress.  The total timing error is given by the following expression: 
\beq
\sigma^2_t = \sigma^2_{TW}+\sigma^2_J + \sigma^2_L + \sigma^2_{TDC},
\label{eq:time-error}
\eeq
where $\sigma_{TW}$ is the contribution of the time walk of the input signal, $\sigma_J$ is the contribution of the time jitter, $\sigma_L$ is the noise contribution by Landau fluctuations and  $\sigma_{TDC}$ is the contribution of the TDC that digitises the timing signal. 
Out of the different components the contribution of the jitter is the most relevant one. At the MIP level it is given by: 
\beq
\sigma_J = \sigma_{Noise}/(dV/dt).
\label{eq:jitter}
\eeq
From Eq.~\ref{eq:jitter} it can be seen that the jitter is controlled by two components One is the electronics noise $\sigma_{Noise}$ generated mainly by the pre-amplifier of the readout circuit. The second component is the slew rate (slope), $dV/dt$. It is obvious that a quick rise of the pulse is beneficial for a good time resolution. On the other hand also a large signal reduces the jitter. For N concurrent MIP the overall jitter reads thus:
\beq
\sigma_J(N) = (1/N) \cdot \sigma_J(MIP).
\label{eq:jittern}
\eeq
This observation is the deeper reason why for example in the results of the beam test measurement of CMS, the best time resolution of about 20\,ps is achieved for large signals, see Fig.~\ref{fig:time-resol}.

An alternative to conventional (passive) Si diodes are LGAD sensors where LGAD stands for {\em Low Gain Avalanche Detectors}. In this case an extra doping layer is added to the usual diode. By this, locally electric fields of 300\,kV/cm are achieved that are sufficient to start an avalanche of electrons and holes. Since electrons are  absorbed immediately it is the hole drift that dominates the actual slope. The gain current is also a function of the weighting field $1/d$ resulting in the following basic relation between time slew $dV/dt$, gain $G$ and sensor thickness $d$

\beq
\frac{dV}{dt} \propto \frac{G}{d}
\label{eq:gain-lgad}
\eeq

From this it is obvious that an optimal layout minimises the sensor thickness and maximises the gain. For the latter however a significant optimisation process is needed. High gain may entail a number of drawbacks such as higher noise (in general the noise increases faster than the signal) and higher leakage current. The presence of high currents may also result in a higher thermal load of the detectors. The general observation is however that LGAD sensors are superior to the conventional silicon sensors in particular for small amplitudes. A factor of 3, i.e. ~70\,ps compared with 200\,ps, at the MIP level has been reported at the workshop. The coming years will see a sustained R\&D programme for the development of ultra fast silicon devices. The goal is to find optimal signal over noise working points assuring at the same time a high gain. It was found that high gains may require shaping times as low as 1\,ns. The importance of a large signal is also exploited in the APD approach where time resolutions as good as 10\,ps at high amplitudes are reported. The ongoing R\&D programme makes heavy use of the previously introduced simulation programme Weightfield. 

\subsection{Timing measurements - Signal shaping}

\begin{figure}[h]
  {\centering
    \includegraphics[width=0.9\textwidth]{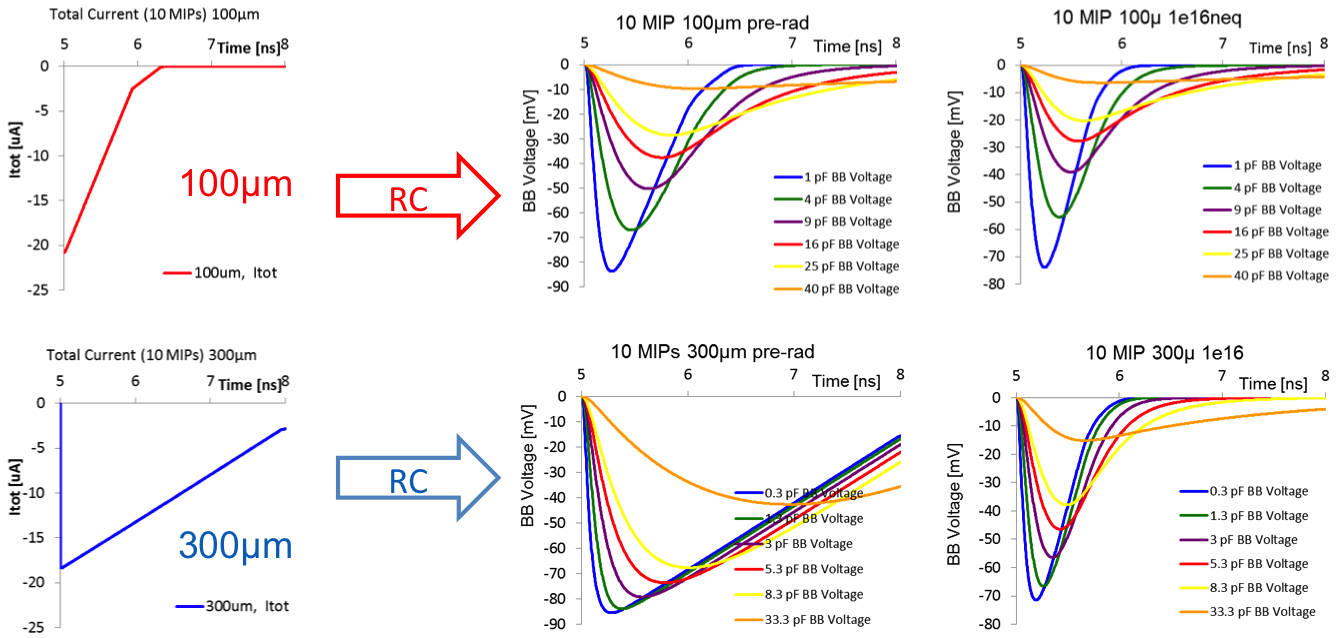}
    \caption{\sl  Schematic view of signal formation and subsequent shaping by an $RC$ circuit emulating a broadband amplifier for silicon sensors of a thickness of 100 and 300$\mu m$, respectively. The right column shows the modification of a signal after irradiation with a dose of  $10^{16}$\,neq. The distributions were obtained with the Weightfield simulation.}
    \label{fig:weightfield-sim}
  }
\end{figure}

A few basics of signal shaping and radiation effects can be extracted from Fig.~\ref{fig:weightfield-sim} obtained with the Weightfield simulation. 

The simulation starts with an energy deposition of 10 MIPs. The immediate rise of the current is followed by linear decrease that is longer for thicker wafers. The pre-amplification, assumed to be realised by a broadband amplifier with 50\,$\Omega$ input impedance, dilutes the time slew signal from  $dV/dt = \infty$ to a finite value for $dV/dt$. The time slew varies strongly with the overall capacitance of the Si counter but less with the actual thickness. The integrated signal is however larger for the thicker sensor which is less relevant for time measurements. It is further observed that radiation effects have little influence of the time slew but a large influence on the integrated, i.e. total, signal. From this study and the previous considerations outlined for the LGAD type sensors it is obvious that the read out electronics play a crucial role in the success of time measurements with Si devices. 

\subsection{Timing measurements - Clock distribution}
\begin{figure}[h]
  {\centering
    \includegraphics[width=0.9\textwidth]{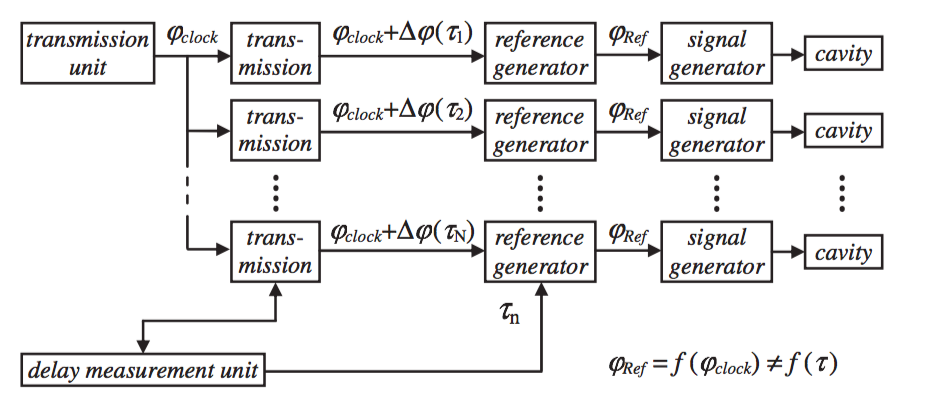}
    \caption{\sl  Basic principle of the clock distribution system described in the text.}
    \label{fig:clock-basics}
  }
\end{figure}

Precise timing requires an accurate clock distribution across a big detecting system. At the workshop an option has been discussed that is based on a timing system developed for the synchronisation of accelerator devices at the FAIR facility~\cite{bib:bousonville}. A time signal to control the ramp of accelerating cavities is transmitted to various places at the accelerator complex that are spatially several hundred meters apart. The system  consists  basically of a clock-transmitting unit, a measurement unit and a reference generator. The basic principle is shown in Fig.~\ref{fig:clock-basics}.

These three units share the {\em same} optical fibre network by means of the application of the {\em Dense Wavelength Division Multiplex Method} (DWDM). Therefore delays between different positions around the accelerator complex can be determined without systematic effects that may be due to different networks for signal transmission and delay control. The measurement unit sends phase corrections to the reference unit such that cavity ramping can happen synchronised around the accelerator complex. With the test setup discussed in Ref.~\cite{bib:bousonville} phase synchronisation with a jitter lower than 1\,ps seem to be achievable. The described system is adopted by the LHC experiments CMS and TOTEM and the characterisation of the clock transmission chain was presented at the workshop. The system for CMS/TOTEM~\cite{bib:cern-totem} is based on a TICDCM6208  chip as clock source that features a particularly small jitter of 1\,ps and better.

\section{Front end electronics for Si sensors}

\begin{figure}[h]
  {\centering
    \includegraphics[width=0.2\textwidth]{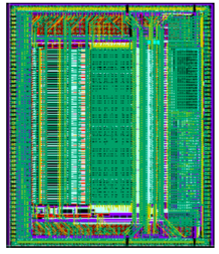}\hspace{2.0cm}
    \includegraphics[width=0.3\textwidth]{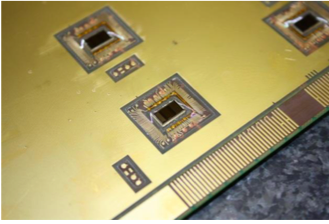}
    \caption{\sl  Left: Picture of SKIROC2 ASIC for CALICE silicon tungsten electromagnetic calorimeter. Right: SKIROC2 ASICs wire-bonded on PCB.}
    \label{fig:skiroc}
  }
\end{figure}

Front end electronics for highly granular calorimeters are developed for more than a decade. So far the needs were mainly pushed for the needs of detectors at linear colliders. Most of the work is carried out within the CALICE Collaboration. The ASICs that are currently under study are SPIROC, HARDROC and MICROROC for hadronic calorimeters and SKIROC for electromagnetic calorimeters with silicon, see Fig.~\ref{fig:skiroc} 

All ASICs are based on SiGe 350\,nm technology and integrate signal amplification, signal shaping and digitisation within one ASIC. The ASICs are  auto-triggering meaning that only in case of sufficiently high energy depositions the signals are propagated to the data acquisition. The goal is to lower trigger thresholds to values as low as 1/2\,MIP corresponding to a charge of 2\,fC in high resistive p-i-n $325\,{\rm \mu m}$ thick diodes as currently used by the CALICE prototype. This goal is met by the SKIROC ASIC. The mentioned ASICs can be operated in power pulsed mode. Power pulsing implies a periodical shut-down and relaunch of the bias currents that polarise the various stages of the ASIC. The typical operation mode is given by the time structure of the beams of the International Linear Collider where the ASICs will be enabled during the 2\,ms duration of a bunch train and disabled during the about 198\,ms between the bunch trains. Power pulsing avoids the need of large cooling systems for the detector supporting thus greatly the design goal of compact and highly granular detectors. Power pulsing will reduce the power consumption of the ASICs to values as low as 25\,$\mu W$/channel. In general all developed ASICs for granular calorimeters show satisfactory behaviour in beam tests and on test benches which is particularly true for the operation in power pulsed mode. In the latter case a switch on-time of a few 100\,$\mu s$ will have to be taken into account during the detector operation. 
The ASICs developed for CALICE and in particular the SKIROC ASICs constitute the ideal starting point for the development of ASICs for the R\&D for the detector HGCAL project of CMS and the HGTD project of ATLAS~\cite{bib:cern-atlas}. The first "LHC variant" that is available is the SKIROC\_CMS ASIC. The 40\,MHz circular buffer and the short shaping time of 25\,ns meet already the requirements of the current R\&D phase. The ASICs feature a dual charge amplifier meaning that they can read out n-type and p-type wafers at the same time. 

Loosely speaking, forward calorimeters at linear collider detectors put similar requirements on the front end electronics as the LHC detectors. The FLAME ASIC, a 16 channel ASIC in CMOS 130\,nm,  is developed by the FCAL Collaboration~\cite{bib:url-fcal} to meet these requirements.

\subsection{Aspects for timing measurements}  

As already mentioned earlier the electronics jitter plays a crucial role for an excellent time resolution. Taking into account the elements of the readout circuit the jitter can be expressed as follows:
\beq
\sigma_J = \alpha \frac{C_d}{\sqrt{g_m}} \frac{\sqrt{t^2_{r\_a} + t^2_d } }   {\sqrt{t_{r\_a}}}.
\label{eq:jitter-elec}
\eeq
Here $g_m$ is the transconductance of the pre-amplifier, $t_d$ is the signal duration and $t_{r\_a}$ is the time constant of the output of a pre-amplifer that characterises its bandwidth. Further $C_d$ is the detector capacitance  and $\alpha$ is a proportional constant. From Eq.~\ref{eq:jitter-elec} it is obvious that the detector capacitance should be as small as possible and the preamplifier has to assure a big transconductance.  The bandwidth of the pre-amplifier has to be carefully adjusted to the detector layout since the jitter in Eq.~\ref{eq:jitter-elec} is optimal if $t_d = t_r$.  Further aspects that need to be taken into account for an optimal time measurements is the time of arrival at a given threshold that may depend on the input signal. This is represented by $\sigma_{TW}$ in Eq.~\ref{eq:time-error}.
A given threshold may be crossed at different times since the form of a signal changes  with its magnitude. Typical differences are of the order of 100ps/MIP for $Q>2$\,MIP. The situation worsens towards small signals. 

For the optimisation of the aforementioned issues several broadband pre-amplifier configurations, transimpedance (TZ)  and common emitter (CE), as well as ASIC technologies (65/130\,nm) are currently investigated. 

%It has to be reminded at the beginning that the signal that is input to the front end electronics is determined by some characteristic features of the Si sensor. A sensor can be approximated by a capacitance $C_d$. Parallel to that the input is terminated by a typically 50 Ohm resistance. The input signal to the front end electronics is now given by the time constant $\tau = C_d R_s$. For a small signal duration the voltage that is input from the pre-amplifier 

\section{Mechanics - Housing and Cooling}
\begin{figure}[h]
  {\centering
    \includegraphics[width=0.25\textwidth]{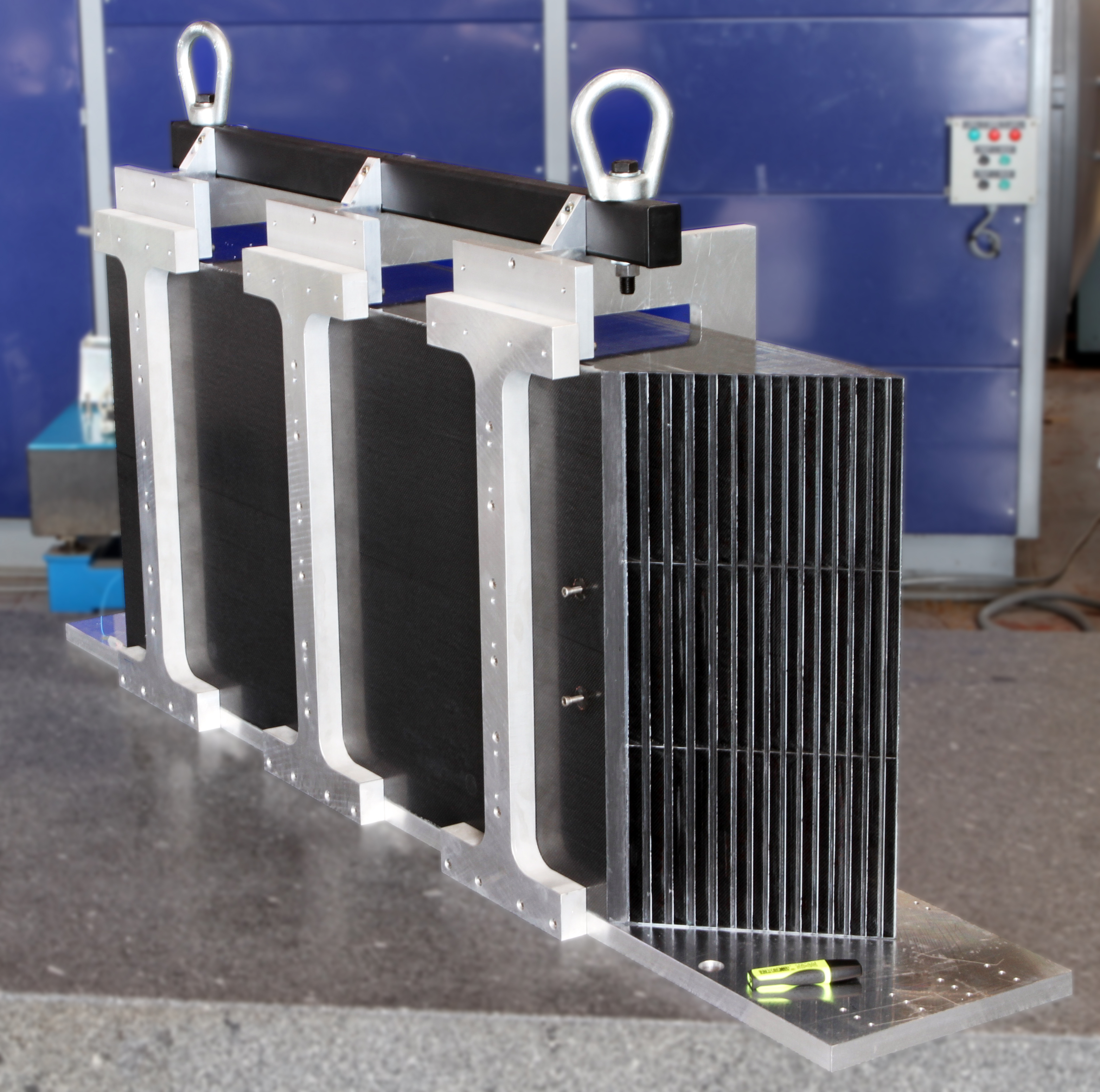}
    \hspace{0.5cm}
    \includegraphics[width=0.25\textwidth]{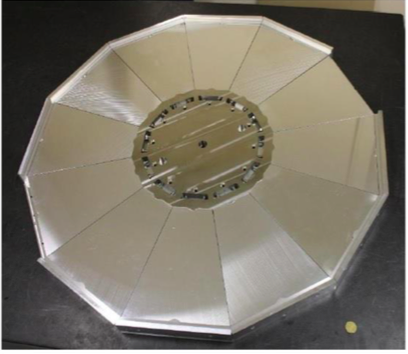}
     \hspace{0.5cm}
    \includegraphics[width=0.25\textwidth]{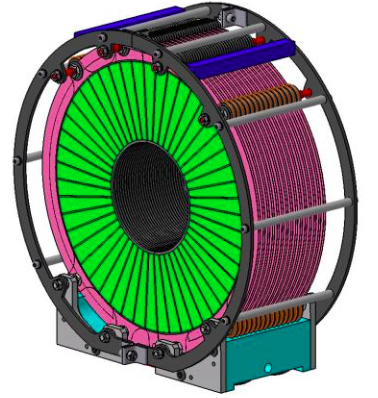}
    \caption{\sl  Left: Alveolar structure for the CALICE silicon tungsten electromagnetic calorimeter prototype. Middle: Mould for alveolar disk structures prototypes for CMS HGCAL. Right: Schematic view of luminosity calorimeter for lineat collider detectors.}
    \label{fig:calice-tech-proto}
  }
\end{figure}

The mechanics design has to support the ultra-compactness of the sensitive parts of the detector. For detectors at future electron-positron colliders high granular electromagnetic silicon tungsten calorimeters are a baseline solution for the central detector part which is typically divided into a barrel and an end-cap part.
The CALICE collaboration has constructed so called alveolar structures. An example for such a structure for the barrel part of the ILD Detector~\cite{Behnke:2013lya} is given in the left part of Fig.~\ref{fig:calice-tech-proto}. 

The structure is realised by a tungsten-carbon reinforced epoxy (CRP) and serves as the absorption medium while ensuring at the same time the mechanical stability of the detector. The dimensions of the this prototype are $1560\time 186 \times 550\,{\rm mm^3}$. 
The LHC experiments consider to use highly granular calorimeters rather in the forward parts, which motivates a different mechanical layout. The fabrication of first test modules for CMS took example from the CALICE structures, however instead of moulding rectangular alveoli the cassettes feature a petal like shape to meet the specification of the forward region of the CMS detector, see middle part of Fig.~\ref{fig:calice-tech-proto}. At the time of the workshop ATLAS considered using rectangular alveoli as CALICE, however, which a somewhat shorter length. Finally, the right part of Fig.~\ref{fig:calice-tech-proto} shows a schematic view of the luminosity calorimeter for linear collider detectors.

Cooling is a important aspect of the detector mechanics. The compact design of the detectors leaves only very little space for the cooling system. The design of the power pulsed front end electronics for the ILC detectors facilitates the design of the cooling system as virtually no heat will be produced by the front end electronics inside the detector volume. Residual heat can be transported to well defined spots at detector layer extremities that are connected to a cooling system. This clearly supports the compactness of the final system. 

The harsh radiation environment at the LHC requires that the entire detector is operated in a cryostat and cooled down to $-30^{\circ}{\rm C}$. Leakage currents and front electronics operated continuously as is the case at the LHC experiments but also for other proposals for future detectors require that full cooling has to be integrated into the detector volume. This may be realised by micro-channel cooling. Here, small pipes are embedded into the tungsten plates. These pipes are flushed by $CO_2$ as the cooling agent. 

\section{Vendors}

\begin{figure}[h]
  {\centering
    \includegraphics[width=0.6\textwidth]{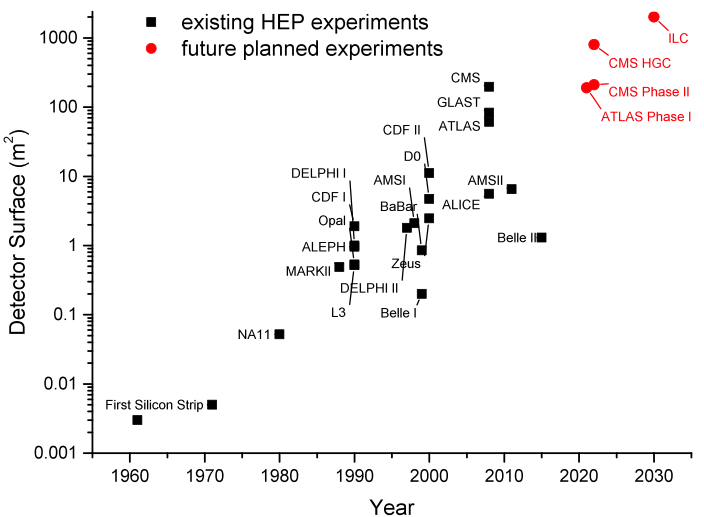}
    \caption{\sl  Surface covered by silicon sensors in past, present and future particle physics experiments.}
    \label{fig:si-needs}
  }
\end{figure}

The ever increasing surface covered by silicon in particle physics experiments is illustrated in Fig.~\ref{fig:si-needs}.  

Major calorimeter systems will comprise large areas of silicon reaching a surface of up to $2500\,{\rm m^2}$ in case of Linear Collider Detectors, The CMS HGCAL will cover around $600\,{\rm m^2}$. These considerable quantities needed already in the next 5-10 years requires the establishment of a network of vendors capable of producing large quantities at a reproducible high quality {\em and at a reasonable price}. 
Prior to the workshop a market survey for CMS/ATLAS tracker needs has been launched that syncs in also the needs for the CMS HGCAL project. Eight companies in Europe, Asia (Japan) and the US have replied to the survey. The result of the survey is summarised in the AIDA-2020 Note~\cite{Bergauer:2238327}.  The survey is an adequate step to address the various producers as one particle physics community instead of addressing them as separate groups or projects. 
Today it is still fair to say that  currently there is only one vendor worldwide able to satisfy the requests, Hamamatsu Photonics. The workshop reported on the effort to gain Infineon (Germany) as a second large scale producer. The design of the detectors may need to be adapted to production lines that will be used by the vendors. Currently there is discussion ongoing whether the silicon wafers (square or hexagonal) will be fabricated from 6" or 8" technology. While currently the 6" inch is standard the tendency goes toward 8" and it is advisable  that each future project has a strategy for the 8" technology.

%It might be preferable to have "generic" silicon producers among the vendors, rather than silicon detector manufacturers. Such "generic" producers are generally more hesitant to perform an R\&D program without the real prospect of a full detector production.     
%With the upgrade programme of the LHC this situation has changed considerably.  

So far the discussion assumed diodes as considered for tracker upgrades, HGCAL (CMS) or Linear Collider experiments. As mentioned above for the optimisation of the timing measurement LGAD's may be the sensor of choice. Also here the step will need to be made from tiny systems as currently under test to full size detector systems covering a few ${\rm m^2}$. The ATLAS collaboration is entertaining  tight contacts with the Spanish company CNM for the HGTD upgrade project.   

\section{Conclusion and outlook}
The summarised workshop brought together, maybe for the first time, main stakeholders of highly granular large scale silicon based calorimeters. The enormous  potential of these detectors in terms of particle separation is demonstrated since more than ten years by the CALICE collaboration. These kind of detectors need very compact designs in terms of readout and mechanics. Both requirements are met with the current generation of ASICs and alveolar structures that are adapted from the CALICE R\&D programme to the needs of the LHC experiments.

The latter add two new crucial aspects to the technology. The first is the radiation hardness that maybe will drive the choice of the silicon wafer type and may have a repercussion on the final layout of the front electronics. The second one is the request for a precision timing. The goal of reaching a precision of the order of 20\,ps or better put new challenges to read out systems in terms of signal shaping but also clock distribution has to be pushed towards new limits.  

The detectors discussed so far can be realised as forward and central detector components until the year 2030, i.e. for the next generation of particle physics experiments. These ultrahigh calorimeters will have about $10^8$ readout cells. Three orders of magnitude more can be reached by basing calorimeters on ultrahigh pixelised CMOS or MAPS ASICs. These detectors are in a comparatively early R\&D phase but are still options for upgrades of the ALICE experiment~\cite{bib:cern-alice} or even as an alternative for Linear Collider Detectors. 

In any case it is extremely important that the projects and the R\&D of the different branches of highly granular calorimeters with silicon will be monitored (perhaps even coordinated). This includes the contact with industrial partners. AIDA-2020 offers the ideal platform for this purpose. 

\section*{Acknowledgements}
The authors would like to thank the management of the AIDA-2020 project for giving us the opportunity to organise the workshop. We would like to thank all speakers for their high quality contributions, see also Appendix~\ref{app:speakers}, and the participants for the fruitful discussions. The seamless organisation of the workshop was made possible by DESY and the secretaries Natalia Potylitsina-Kube, Livia Elena Bacalan and Sabrina El Yacoubi. The workshop has received funding from the European Union�s Horizon 2020 Research and Innovation programme under Grant Agreement no. 654168.

\bibliographystyle{utphys_mod}
\bibliography{siws-sum}
\begin{appendices}
\newpage
\section{List of Speakers} \label{app:speakers}
In the following the list of speakers with the title of the talk and a link to the presentation are given.
\begin{itemize}
\item Marcello Manelli (CERN), {\em "Summary of the state of the art"}, \url{https://indico.cern.ch/event/468478/contributions/2135135/attachments/1289836/1921060/MM_High_Granularity_Si_Calorimeters_Overview_and_State_of_the_Art.pptx};
\item Walter Snoeys (CERN), {\em "Ultrahigh granular calorimeters with HV CMOS and MAPS"}, \url{https://indico.cern.ch/event/468478/contributions/2135136/attachments/1289971/1920682/W_Snoeys_AIDA2020_cal.pdf};
\item Arabella Martelli (CERN), {\em "Results of beam-tests of HGCAL diodes/sensors for fast-timing measurement"} \url{https://indico.cern.ch/event/468478/contributions/2182179/attachments/1289952/1920650/SiHGC_TBtiming_DesyWorkshop_June2016.pdf};
\item Nicolo Cartiglia (University and INFN Torino), {\em "Signal formation and timing in LGAD sensors"}, \url{https://indico.cern.ch/event/468478/contributions/2135146/attachments/1290321/1921375/LGA_Signal_Cartiglia.pdf};
\item Hartmut Sadrozinski (UCSC), {\em  Signal formation and timing in conventional P-I-N Diodes"}, \url{https://indico.cern.ch/event/468478/contributions/2135147/attachments/1290317/1921371/HFWS-Cal-Si.pdf};
\item Michael Moll (CERN), {\em "Radiation hardness issues for silicon sensors"}, \url{https://indico.cern.ch/event/468478/contributions/2135148/attachments/1290271/1921271/1600613-RD50-AIDA2020.pdf};
\item Thomas Bergauer (\"OAW), {\em "Issues of Si Wafer production and overview of tentative Si producers"}, \url{https://indico.cern.ch/event/468478/contributions/2135149/attachments/1290245/1921216/Si-Production_Bergauer_AIDA-2020.pdf};
\item Christophe de la Taille (CNRS/IN2P3/OMEGA), {\em "Front end electronics"}, \url{https://indico.cern.ch/event/468478/contributions/2135165/attachments/1290561/1921884/CdLT_HGSi_14jun16.pdf};
\item Michele Quinto (CERN), {\em "Digital readout and clock distribution"}, \url{https://indico.cern.ch/event/468478/contributions/2135166/attachments/1290587/1921986/digital_readout_and_clock_distribution.pdf};
\item Denis Grondin (CNRS/IN2P3/LPSC), {\em "Mechanical structures"}, \url{https://indico.cern.ch/event/468478/contributions/2135172/attachments/1290585/1922341/MECHANICAL_STRUCTURES.pdf};
\item Marek Idzik (AGH-UST), {\em "Compact forward calorimetry at future linear colliders"} \url{https://indico.cern.ch/event/468478/contributions/2135173/attachments/1290631/1922300/idzik_aida2020_SiWorkshop_2016_06.pdf};
\item Roman P\"oschl (CNRS/IN2P3/LAL), {\em "Summary and next steps"}, \url{https://indico.cern.ch/event/468478/contributions/2135180/attachments/1290786/1922442/talk140616.pdf}.
\end{itemize}

\end{appendices}
\end{document}